\documentclass[aps,pra,epsfigure,twocolumn,superscriptaddress]{revtex4-2}
\usepackage[colorlinks=true,linkcolor=blue,urlcolor=blue,citecolor=blue,pdfusetitle]{hyperref}
\usepackage[utf8]{inputenc}
\usepackage[english]{babel}
\usepackage{amsmath}
\usepackage[caption = false]{subfig}
\usepackage{blindtext}
\usepackage[table,xcdraw]{xcolor}
\usepackage{lipsum}
\usepackage{amsfonts}
\usepackage{bbold}
\usepackage{bbm}
\usepackage{amssymb}
\usepackage{enumerate}
\usepackage{color}
\usepackage{xcolor}
\usepackage{latexsym}
\usepackage{physics}
\usepackage{braket}
\usepackage{times,txfonts}
\usepackage{bm}
\usepackage{graphicx,epstopdf}

\usepackage[cal=boondox]{mathalpha}

\usepackage{orcidlink}

\begin{document}

\title{Light Statistics from Large Ensembles of Independent Two-level Emitters: Classical or Non-classical?}

\author{M. Bojer}
\email{manuelbojer6@gmail.com}
\affiliation{Friedrich-Alexander-Universität Erlangen-Nürnberg, Quantum Optics and Quantum Information, Staudtstr. 1, 91058 Erlangen, Germany}
\author{A. Cidrim}
\affiliation{Departamento de F\'isica, Universidade Federal de S\~ao Carlos,
	Rodovia Washington Lu\'is, km 235—SP-310, 13565-905 S\~ao Carlos, SP, Brazil}
\author{P. P. Abrantes}
\affiliation{Departamento de F\'isica, Universidade Federal de S\~ao Carlos,
	Rodovia Washington Lu\'is, km 235—SP-310, 13565-905 S\~ao Carlos, SP, Brazil}
\affiliation{Instituto de F\'isica, Universidade Federal do Rio de Janeiro, 21941-972 Rio de Janeiro, RJ, Brazil}
\author{R. Bachelard}
\affiliation{Departamento de F\'isica, Universidade Federal de S\~ao Carlos,
	Rodovia Washington Lu\'is, km 235—SP-310, 13565-905 S\~ao Carlos, SP, Brazil}
\author{J. von Zanthier}
\affiliation{Friedrich-Alexander-Universität Erlangen-Nürnberg, Quantum Optics and Quantum Information, Staudtstr. 1, 91058 Erlangen, Germany}

\date{\today}

\begin{abstract}
    We investigate the photon statistics of an ensemble of coherently driven non-interacting two-level atoms in the weak driving regime. As it turns out, the system displays unique emission characteristics that are strongly in contrast to the emission of classical oscillating dipoles. By deriving the second-order autocorrelation function, we show that extraordinary two-photon correlations are obtained, ranging from strong antibunching to superbunching. These features are enhanced by disorder in the emitter positions, and the control parameter is the number of excitations in the system. We observe the appearance of bunching and antibunching when the light is scattered by the atoms predominantly coherently, i.e., mimicking classical Rayleigh scattering, whereas thermal photon statistics is obtained when the light is scattered via spontaneous decay, a well-known quantum effect. The underlying mechanism is the interplay between coherent scattering, which exhibits spatial fluctuations due to interference, and dissipation in the form of isotropic spontaneous decay.
\end{abstract}

\maketitle


{\it Introduction---}The scattering of light by an ensemble of particles is a long-standing problem in physics, starting from classical descriptions employing Maxwell's equations to the first quantum mechanical treatment by Einstein, tackling for the first time stimulated emission by two-level atoms~\cite{1917PhyZ...18..121E}. Later, Mollow, and then Carmichael and Walls, studied the emission spectra of the scattered light, deriving the Mollow triplet as well as a coherent delta-peak in a fully quantum mechanical treatment~\cite{Mollow1969,Carmichael1976}. The subsequent investigation by Glauber of the photon statistics of the scattered light divided light sources into three categories, characterized by antibunched, uncorrelated, and bunched photon arrival times. In particular, single-photon (quantum) emitters such as atoms~\cite{Clauser1974,Kimble1977}, ions~\cite{Diedrich1987}, or color centers~\cite{Kurtsiefer2000} display antibunching of the scattered photons, while (classical) coherent or thermal sources exhibit uncorrelated or bunched photon emission, respectively. 

From a classical perspective, the scattering of incident coherent monochromatic light by an ensemble of scatterers can be described by the Lorentz oscillator model, which treats the emitters as classical damped driven harmonic oscillators. The quantum mechanical counterpart of this approach involves a density matrix $\hat{\rho}$, whose temporal evolution is determined by a master equation. The latter reads in the interaction picture $\dot{\hat{\rho}} = [\hat{\rho}, \Omega (\hat{b}^\dagger - \hat{b})] - \gamma (\hat{b}^\dagger \hat{b} \hat{\rho} + \hat{\rho} \hat{b}^\dagger \hat{b} - 2 \hat{b} \hat{\rho} \hat{b}^\dagger)$, with $\Omega$ being the Rabi frequency of the resonant pump (proportional to its amplitude), $\hat{b}^\dagger$ ($\hat{b}$) the rising (lowering) operator of the harmonic oscillator excitations, and $\gamma$ the damping rate. This dynamics admits as steady state solution the coherent state $\ket{\alpha}$, i.e., $\hat{\rho}_{\mathrm{ss}}=\ket{\alpha}\bra{\alpha}$, whose amplitude is given by $\alpha = -\Omega/\gamma$~\cite{SM}. If the particles do not interact with each other and are all driven with the same laser intensity, the state of the ensemble is simply the $N$th tensor product of the single-particle state, $\hat{\rho}^{(N)} = \otimes_{\mu=1}^{N} \hat{\rho}_{ss}^\mu$. Since the field scattered by the oscillators is given by $\hat{E}^{(+)} \sim \sum_{\mu=1}^Ne^{\phi_{\mu}}\hat{b}_\mu $, with $\phi_{\mu}$ a position-dependent phase, and the coherent state is an eigenstate of $\hat{b}$ (that is, $\hat{b}_\mu \ket{\alpha_\mu}=\alpha_\mu\ket{\alpha_\mu}$), then the coherent nature of the oscillators $\hat{\rho}_{\mathrm{ss}}$ is transferred to that of the light, so that all the moments of the light display values expected by a coherent state. In physical terms, this implies that the radiation from $N$ independent damped driven quantum harmonic oscillators is a coherent state for the light, independently of the particle number $N$. In other words, the coherence properties of the driving laser light are refound in the scattered light. Concerning the photon statistics, we therefore expect an uncorrelated emission of photons, corresponding to a spatially isotropic and temporally flat second-order correlation function.

Equally, addressing the $N$ atoms as two-level emitters, it has been shown that for a weak drive (linear optics regime), and thus considering only the ground state and the single-excitation manifold of the $N$ atoms, the equations of motion of the dipoles of the atoms can be described by a classical dipole model, fully consistent with Maxwell's equations~\cite{Svidzinsky2010,Cottier2018}. Correspondingly, by expanding the $N$-atomic  steady state $\hat{\rho}^{(N)}$ to the first order in the saturation parameter $s$, a one-to-one mapping between  $\hat{\rho}^{(N)}$ and the coherent state $\otimes_{\mu=1}^{N}\ket{\alpha_\mu}\bra{\alpha_\mu}$ introduced above can be achieved~\cite{SM}. These results suggest again that in the scattered light a photon statistics corresponding to a coherent  source should be observed in the weak drive regime.

On the other hand, a single two-level emitter is known to produce very different light statistics. In particular, its inability to absorb a second photon once in the excited state leads to photon antibunching of the scattered light~\cite{Kimble1977, Diedrich1987}. At the same time, increasing the number of independent emitters can along with the presence of decoherence mechanisms~\cite{2023_Lassegues_Hugbart_pra} lead to the configuration of so-called chaotic light and thus to the emergence of thermal light statistics for the radiated field, characterized by photon bunching. Its statistics can be derived again from a classical description~\cite{Loudon:book}.

\begin{figure*}[t!]
\centering
\includegraphics[width=\textwidth]{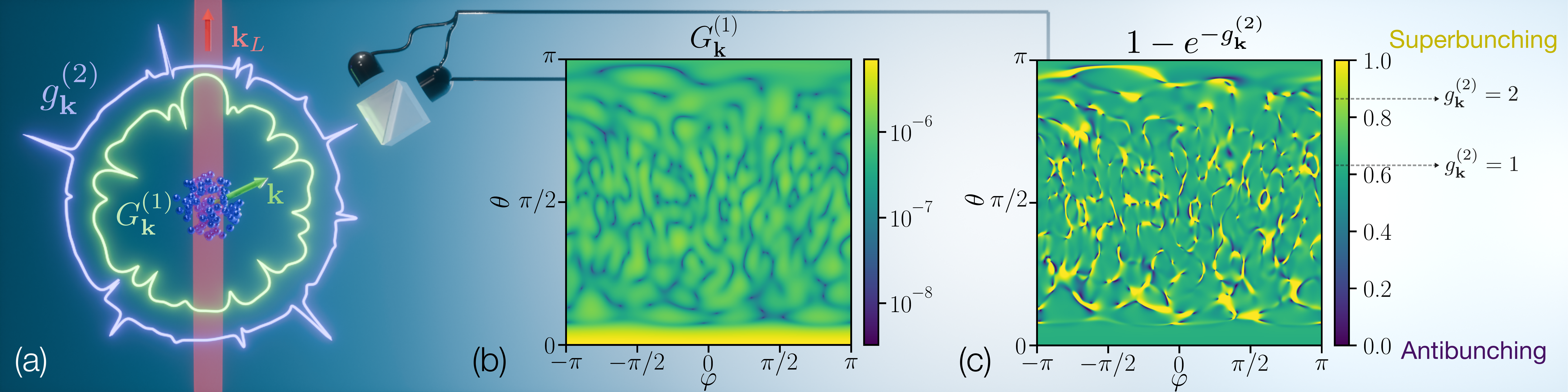}
 \caption{Second-order autocorrelation function for a disordered cloud of $N=100$ atoms and a saturation parameter of $s=10^{-6}$. (a) Atomic configuration together with a cut of the intensity (green curve) and the second-order autocorrelation function (purple curve) in the $xz$-plane. From this typical radiation pattern one can clearly see that the bunching peaks are correlated with the minima of intensity (destructive interference), reflecting the anticorrelation between $g^{(2)}_\mathbf{k}$ and $G^{(1)}_\mathbf{k}$  (Pearson coefficient $r\approx -0.31$, see discussion in the main text). (b-c) Full angular maps of the intensity and the second-order autocorrelation function, respectively. Plotting $1-\exp(-g^{(2)}_{\mathbf{k}})$ allows one to better appreciate its spatial fluctuations. The intensity fluctuates randomly on a scale set by the system size, as expected from speckle theory, and the second-order autocorrelation function shows a similar behavior, with several regions of superbunching and antibunching. The anticorrelation between $G^{(1)}_{\mathbf{k}}$ and $g^{(2)}_{\mathbf{k}}$ can also be appreciated in these plots.}
\label{Fig:sketchy_fig}
\end{figure*}

The question thus arises: Which photon statistics is expected from a weakly driven ensemble of independent two-level emitters for which the only decoherence mechanism is spontaneous emission? In particular, it might be asked whether the light statistics can be inferred from the linear optics regime or whether the small higher-excitation contributions matter. In this letter, we answer this question by deriving explicitly the second-order autocorrelation function of the light scattered by an ensemble of driven non-interacting two-level emitters. In particular, we show that in the weak driving regime extraordinary two-photon correlations are obtained, ranging from strong antibunching to superbunching [see Fig.~\ref{Fig:sketchy_fig}(a)]. Moreover, we demonstrate that, with spontaneous emission taking on the role of decoherence, its interplay with the coherent scattering leads to anomalous autocorrelation functions for the light at all orders. Akin to the case of single emitters whose light statistics can be manipulated by frequency-filtering the radiated light~\cite{LopezCarreno2018,Hanschke2020,Phillips2020}, interference appears here as a key mechanism to shape the photon correlations in large ensembles of quantum emitters.

{\it Two-photon correlations from ensembles of driven two-level emitters---}Let us consider a cloud of $N$ two-level atoms at fixed positions $\textbf{R}_{\mu}$, with ground state $\ket{g}$ and excited state $\ket{e}$, and transition wavelength $2\pi/k$. Let us assume further that the electric dipole transition is driven resonantly by a monochromatic plane wave with wave vector $\textbf{k}_L$, whose intensity $I$ is characterized by the saturation parameter $s=I/I_\textrm{sat}$, with $I_\textrm{sat}$ the saturation intensity of the atomic transition.  Assuming that the emitters are distant enough from each other so that interactions between them can be neglected, the stationary atomic state is a product state $\hat{\rho} = \otimes_{\mu=1}^{N} \hat{\rho}_\mu$, where the single-atom density matrix, obtained from the optical Bloch equations, reads: $\rho_\mu^{ee}=1-\rho_\mu^{gg}=s/2(1+s)$ and $\rho_\mu^{eg}=(\rho_\mu^{ge})^*=-\sqrt{s}/\sqrt{2}(1+s)$. In the far-field limit, the normalized radiation of the emitters is given by the positive frequency part of the scattered electric field operator  $\hat{E}^{(+)}_\mathbf{k} = \sum_{\mu=1}^{N} e^{-i \mathbf{k}\cdot\textbf{R}_\mu} \hat{\sigma}_\mu^-$, with $\hat{\sigma}_\mu^-$ the lowering operator of the $\mu$th emitter and $\mathbf{k}\equiv \mathbf{k}_\text{obs}-\mathbf{k}_L$, with $\mathbf{k}_\text{obs}$ the observation direction. With this operator at hand, one can compute explicitly the normalized equal-time second-order photon autocorrelation function for independent emitters~\cite{SM,2023_Lassegues_Hugbart_pra}:
\begin{widetext}
\begin{equation}\label{Eq:g2}
    g^{(2)}_\mathbf{k} = \frac{\braket{\hat{E}^{(-)}_\mathbf{k} \hat{E}^{(-)}_\mathbf{k} \hat{E}^{(+)}_\mathbf{k} \hat{E}^{(+)}_\mathbf{k}}}{\braket{\hat{E}^{(-)}_\mathbf{k}\hat{E}^{(+)}_\mathbf{k}}^2}	
    = \frac{2sN\left[2 +s(N-1)\right]+ 4s(N-2)\left|S(\mathbf{k})\right|^2 + \left|S^2(\mathbf{k}) - S(2\mathbf{k})\right|^2}{\left(sN + |S(\mathbf{k})|^2\right)^2}\,,
\end{equation}
\end{widetext}
where we have introduced the structure factor $S(\mathbf{k}) = \sum_{\mu=1}^{N} e^{i\mathbf{k}.\mathbf{R}_\mu}$. Note that in the denominator of Eq.~(\ref{Eq:g2}) one can identify the (normalized) intensity
\begin{equation}\label{Eq:G1}
G^{(1)}_\mathbf{k}= \braket{\hat{E}^{(-)}_\mathbf{k}\hat{E}^{(+)}_\mathbf{k}} \propto sN + |S(\mathbf{k})|^2\,,    
\end{equation}
where $sN$ is the contribution of the $N$ emitters to the intensity due to spontaneous emission, and  $|S(\mathbf{k})|^2$ the one due to coherent scattering. 


{\em Photon-photon correlations in ordered arrays---}Atomic systems trapped in optical lattices~\cite{Jordens2008} or tweezers~\cite{Barredo2016,Endres2016} allow to produce periodic arrays of quantum emitters. In these systems, the interactions between the atoms can be used to control the flow of light beams~\cite{Rui2020,Srakaew2023}, addressing selectively the one- and two-photon component~\cite{Zhang2022,Pedersen2023}. Differently, we here show that extraordinary two-photon correlations can already be obtained using independent non-interacting emitters.

Let us start with a regular arrangement of $N$ atoms, for which the well-defined differences in optical path between the emitters allows us to investigate the interference phenomena and light statistics analytically. For a one-dimensional (1D) chain with spacing $d$ along direction $\hat{\mathbf{n}}$, the structure factor reads $S(\mathbf{k}) =(1-e^{iN\phi})/(e^{-i\phi}-1)$, where $\phi= d\,\hat{\mathbf{n}}.\mathbf{k}$ is a geometric phase~\cite{SM}. The $S(\mathbf{k})$ term, proportional to the coherent field, exhibits in this case an oscillatory behavior typical of the interference pattern expected from a periodic arrangement. Focusing on the destructive interference condition, $S(\mathbf{k})=0$ leads to $S(2\mathbf{k})=0$ [unless $\phi=(2m+1)\pi$, with $m \in \mathbb{Z}$ and $N$ even, see~\cite{SM} for details], which in turn results in
\begin{equation}
\label{eq:g2ord}
g^{(2)}_\mathbf{k}=\frac{4}{sN}+2-\frac{2}{N}.
\end{equation}
Hence, according to Eq.~\eqref{eq:g2ord}, arbitrarily strong superbunching $g^{(2)}_\mathbf{k}(0)\sim 4/sN$ can be achieved in the limit $sN\to 0$, which scales inversely with the number of excitations $sN$ in the system.

On the other hand, antibunching is achieved under the condition $S^2(\mathbf{k})=S(2\mathbf{k})$, corresponding for periodic systems to $\phi=2m\pi/(N-1)$, with $m\in\mathbb{Z}$. This leads to $S(\mathbf{k})=e^{i\phi}$ and, in the limit $sN\to 0$, $g^{(2)}_\mathbf{k}\approx 8sN$~\cite{SM}. We thus find that even for large particle numbers, arbitrarily strong antibunching can be obtained, scaling linearly with the number of excitations $sN$.   

Note that the generalization of these results to regular arrays of higher dimensions is straightforward. This is due to the fact that the structure factor for these systems factorizes for each additional dimension [i.e., $S(\mathbf{k})=S_xS_yS_z$]. Therefore, both scalings for superbunching and antibunching for 2D and 3D regularly-spaced arrays follow the same reasoning applied to the 1D case above.


{\em Photon-photon correlations in disordered ensembles---}Although regular lattices benefit from the strongest interference features, with fully constructive interference in given directions, we shall now show that the phenomena of super- and antibunching can actually be enhanced by disorder in the emitters' positions. A typical radiation pattern for an ensemble of independent randomly arranged emitters driven by a plane wave propagating along the $z$-direction ($\theta=0$) is shown in Fig.~\ref{Fig:sketchy_fig}(a). The spatial intensity distribution presents in this case also a random structure [see Fig.~\ref{Fig:sketchy_fig}(b)] with angular fluctuations on a scale set by the system size, as expected from speckle theory~\cite{Goodman2020}. Here, for a low saturation parameter and a moderate particle number ($s=10^{-6}$ and $N=100$), $G^{(1)}_\mathbf{k}$ is largely dominated by the coherent component $|S(\mathbf{k})|^2$ and the associated sum of random phases, with spontaneous emission contributing only with a modest intensity background.

\begin{figure}[t!]
\centering
\includegraphics[width=.5\textwidth]{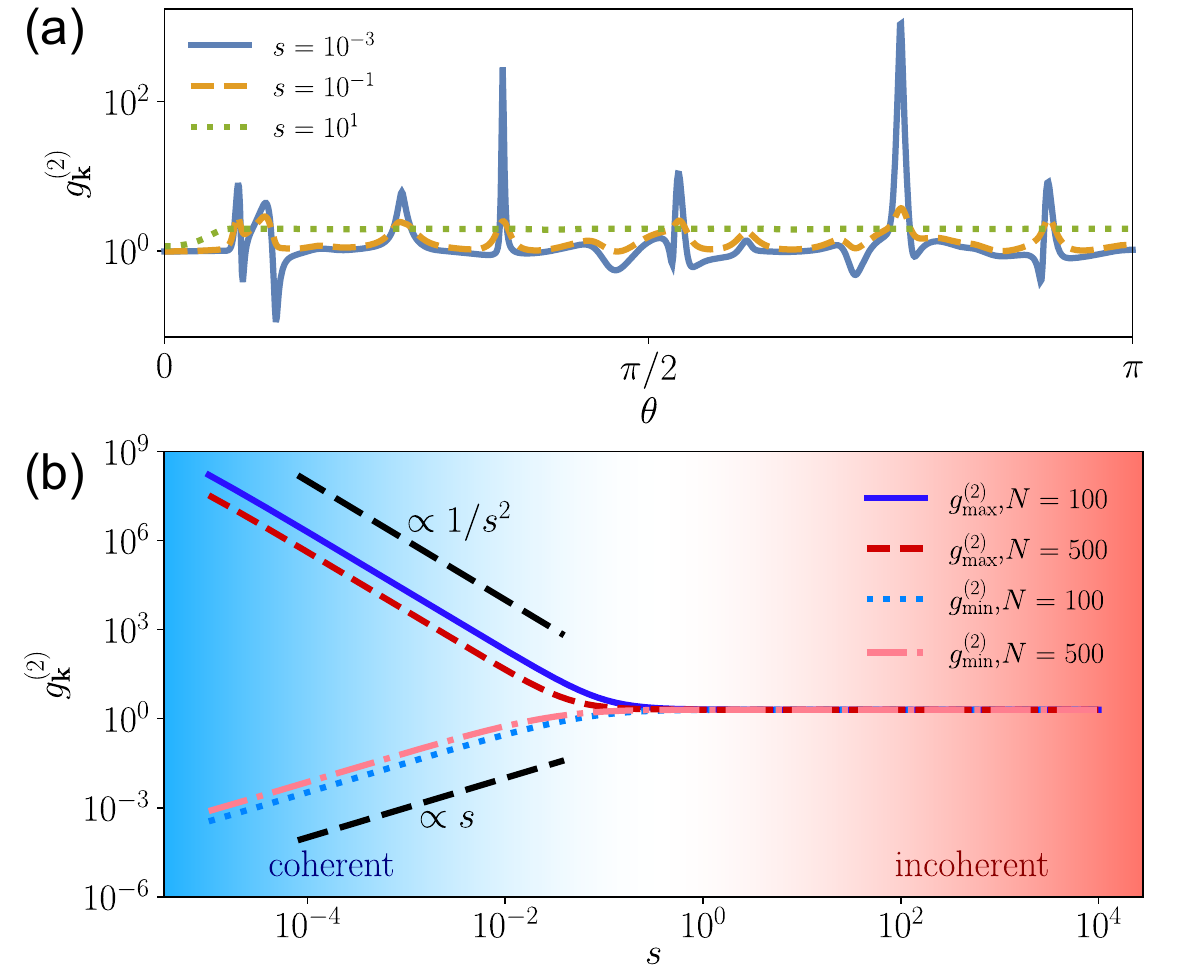}
\caption{(a) Second-order autocorrelation function $g^{(2)}_{\mathbf{k}}$ in the $xz$-plane as a function of $\theta$ for a random cloud of $N=100$ atoms of diameter $k_L r = 6 \pi$, and for different saturation parameters $s$. The amplitude of the fluctuations of the whole pattern increases drastically with $s$ in the limit of vanishing number of excitations, $sN \ll 1$. (b) Scaling of the maximum and minimum of $g^{(2)}_{\mathbf{k}}$ from random ensembles of  $N=100$ and $N=500$ two-level emitters, as a function of the saturation parameter $s$, and averaged over $200$ realizations. In the weak driving limit, the superbunching peaks present a clear scaling $\sim 1/s^2$, whereas the antibunching ones scale linearly in $s$. In addition, both curves show different behaviors as a function of the number of atoms $N$ (see main text). The background gradient highlights that these extraordinary behaviors of the photon statistics occur in the regime where coherent scattering predominates, whereas in the limit of $s\rightarrow \infty$, spontaneous emission takes over and the chaotic light regime is reached.}
\label{Fig:2}
\end{figure}

Interestingly, the second-order photon autocorrelation function $g^{(2)}_\mathbf{k}$ presents a similar random pattern [see Fig.~\ref{Fig:sketchy_fig}(c)], although in anticorrelation with the intensity. The linear correlation between two random variables can be quantified by the ratio $r$ between their covariance normalized by the respective standard deviations --- a quantity also known as the Pearson correlation coefficient, which takes values from $-1$, for ideal anticorrelation, to $+1$, for ideal correlation~\cite{2006_Asuero_Gonzalez_crac}. In our case, the quantities $\log{G^{(1)}_\mathbf{k}}$ and ${1-\exp\big(-g^{(2)}_\mathbf{k}}\big)$ shown in Fig.~\ref{Fig:sketchy_fig} have a coefficient $r\approx-0.31$, indicating a substantial anticorrelation. In particular, one can observe that the peaks of superbunching appear at the minima of the intensity, that is, for destructive interference, as in the ordered case. These directions correspond to the condition $S(\mathbf{k})= 0$, which results in the following normalized second-order autocorrelation function
\begin{align}
	g^{(2)}_\mathbf{k} = \frac{\left|S(2\mathbf{k})\right|^2}{s^2 N^2}+\frac{4}{sN}+2-\frac{2}{N}\,.\label{eq:g2dest}
\end{align} 
Differently from the ordered case with $S(2\mathbf{k})=0$ in directions of destructive interference, the fact that the factor $S(2\mathbf{k})$ does not vanish in these directions in the disordered case leads to a scaling $g^{(2)}_\mathbf{k}\sim \left|S(2\mathbf{k})\right|^2/s^2 N^2$. Estimating the order of magnitude of the $S(2\mathbf{k})$ term, we find that its ensemble average is given by $\left\langle\left|S(2\mathbf{k})\right|^2\right\rangle \sim N$ under the condition of $S(\mathbf{k})=0$~\cite{SM}, so that typically $g^{(2)}_\mathbf{k}\propto 1/s^2 N$. As a consequence, superbunching is expected in the limit of weak drive and in destructive interference directions, enhanced by an additional factor of $1/s$ as compared to the ordered case. This divergence of superbunching for a vanishing drive can be observed in Fig.~\ref{Fig:2}(a), where the angular fluctuations of $g^{(2)}_\mathbf{k}$, for a given atomic configuration, become more extreme as the saturation parameter is reduced.

Note that the $1/s^2$ scaling in the few-excitation limit, $sN \ll 1$, is confirmed by monitoring the maximum value of $g^{(2)}_\mathbf{k}$ over all directions of $\mathbf{k}_{\text{obs}}$ [see Fig.~\ref{Fig:2}(b)]. Here, additionally to $N=100$, we show a second curve with $N=500$ atoms to indicate the different scalings with $N$, whereby both curves represent an average over $200$ realizations. Furthermore, we indicate by the background color the transition from predominantly coherent scattering (blue shaded area) to incoherent scattering (red shaded area) when enlarging the value of the saturation parameter. Contrary to intuition, we here observe the appearance of antibunching when the light is scattered by the atoms predominantly coherently, i.e., mimicking classical Rayleigh scattering, whereas a classical thermal photon statistics is obtained when the light is scattered via spontaneous decay, a well-known quantum effect.

As concerns antibunching, present in the speckled pattern of $g^{(2)}_\mathbf{k}$ in particular directions [see Fig.~\ref{Fig:2}(a)], its scaling is obtained by studying the minima of the numerator in Eq.~\eqref{Eq:g2}. As in the ordered case, it corresponds to the condition $S^2(\mathbf{k})=S(2\mathbf{k})$ in the limit $sN\to 0$. Assuming a large number of emitters $N$, one obtains in this case
\begin{align}
 g^{(2)}_\mathbf{k} \approx 4sN\frac{1+\left|S(\mathbf{k})\right|^2}{\left|S(\mathbf{k})\right|^4}\,.
\end{align}
Note that while the structure factor reaches its maximum value $S(\mathbf{k})=N$ in the forward scattering direction $\mathbf{k}_{\text{obs}}=\mathbf{k}_L$, the here imposed condition $S^2(\mathbf{k})=S(2\mathbf{k})$ cannot be satisfied at the same time since the upper bound for $S(2\mathbf{k})$ is $N$. Hence the condition imposes $|S(\mathbf{k})|\leq \sqrt{N}$, which leads to the lower bound $g^{(2)}_\mathbf{k} \geq 4s$. However, a numerical analysis of the ensemble average of $[1+\left|S(\mathbf{k})\right|^2]/\left|S(\mathbf{k})\right|^4$ shows that its effective scaling is $\left|S(\mathbf{k})\right| \sim N^{1/4}$ rather than $\sqrt{N}$. Therefore, the typical scaling of the second-order autocorrelation function minima is $g^{(2)}_\mathbf{k} \propto 4 s \sqrt{N}$. Consequently, arbitrarily strong antibunching can be achieved when the driving laser power is decreased, although at the cost of a vanishing number of photons. Once more, a more favorable scaling for antibunching with respect to the atom number is obtained in the disordered case as compared to the ordered case, with $g^{(2)}_\mathbf{k}\sim\sqrt{N}s$ instead of $\sim Ns$. Hence, both superbunching and antibunching are enhanced by the disorder in the emitter positions.

Turning now to the strong driving limit, i.e., $s\to \infty$, the $g^{(2)}_\mathbf{k}$ function becomes almost isotropic, with a value close to 2: This is the chaotic light limit, where spontaneous emission dominates the scattering process, readily recovered from Eq.~\eqref{Eq:g2} by setting $s\to\infty$ [see Fig.~\ref{Fig:2}(a)]. Note that in Fig.~\ref{Fig:2}(a), for $s=10$ a substantial deviation from the value 2 is observed only in the $\theta\approx 0$ direction, in which the fully constructive interference leads to the largest coherent component. In this direction of forward scattering, corresponding to $\mathbf{k}_{\text{obs}}=\mathbf{k}_L$ and to a coherent intensity scaling as $N^2$, larger saturation parameters are required to reach the chaotic light regime.


{\it Higher-order extraordinary photon statistics---}  Pairs of photons, signaled by large values of $g^{(2)}_\mathbf{k}$, have become an important resource for quantum information protocols~\cite{Caspani2017}. Even more, bound states of higher number of photons associated with higher-order photon autocorrelations are useful for quantum communication and quantum metrology~\cite{Pan2012}. Such states have been produced, for example, in systems with strong interactions such as those between Rydberg atoms~\cite{Liang2018,Cantu2020} or between emitters and resonators~\cite{Dory2017}. Here we show that the interplay between interference and spontaneous emission allows to manipulate these higher-order photon correlations, with the varying photon statistics observed in $g^{(2)}_\mathbf{k}$ being also present at higher orders.

We start to recall that, for coherent light, the $m$th-order autocorrelation function corresponds to $g^{(m)}_\mathbf{k}\equiv 1$, whereas for chaotic light it reaches the value $m!$. Considering our ensemble of $N$ driven independent two-level emitters, a formula similar to Eq.~\eqref{Eq:g2} can be derived also for higher-order correlation functions~\cite{SM}. This allows us to obtain a condition for generalized ``superbunching", $S(\mathbf{k})=0$, and generalized ``antibunching", $S^{(m)}(\mathbf{k}) = 0$, where we have introduced the following generalized $m$th-order structure factor:
\begin{align}
    S^{(m)}(\mathbf{k}) = \sum_{P_{c_1,...,c_m}} (-1)^{m-\sum_{j=1}^{m} c_j} m! \prod_{l=1}^{m} \frac{S^{c_l}(l \mathbf{k})}{c_l! l^{c_l}}.
\end{align}
Here $P_{c_1,...,c_m}$ denotes an integer partition of $m$ (see Ref.~\cite{SM} for details). In the previously discussed case of $m=2$, there are two integer partitions, namely $P_{2,0}=1+1$ and $P_{0,1}=2$~\cite{SM}. Hence the generalized second-order structure factor is $S^{(2)}(\mathbf{k}) = S^2(\mathbf{k}) - S(2\mathbf{k})$, and the condition $S^2(\mathbf{k})=S(2\mathbf{k})$ for antibunching is recovered.

Moving to higher orders, the following scalings of the $m$th-order photon autocorrelation function with the saturation parameter are obtained:
\begin{align}
\text{superbunching: }& g^{(m)}_{\mathbf{k}} \propto \frac{1}{s^m}\,,\\
    \text{antibunching: }& g^{(m)}_{\mathbf{k}} \propto s\,.
\end{align}
Hence, arbitrarily strong antibunching and superbunching can be achieved, in the vanishing driving limit, at any order of photon-photon correlations. In particular, the condition $g^{(m)}_{\mathbf{k}}<1$ obtained for small values of $s$ shows that the quantum nature of the light produced from the interplay between interference of coherent scattering and spontaneous emission is also displayed by the higher-order photon autocorrelation functions~\cite{2022_Laiho_Marquardt_pla,Laiho2012,Klauder2006}.


{\it Conclusion---}We have demonstrated that coherently driven two-level atoms in the weak driving regime display unique emission characteristics that are in strong contrast to the emission of classical oscillating dipoles. More specifically, strong superbunching (i.e., $g^{(2)}_{\mathbf{k}}\gg 2$) as well as strong antibunching (i.e., $g^{(2)}_{\mathbf{k}}\ll 1$) can be achieved in the limit $sN \ll 1$, where these extreme photon statistics are further enhanced by disorder in the emitter positions. 

The underlying mechanism is the interplay between coherent scattering, which exhibits for an ensemble spatial fluctuations due to interference, and dissipation in the form of spontaneous emission, which is isotropic for independent emitters. This is reminiscent of the light statistics of a driven single quantum emitter, whose antibunching properties actually stem from the interference between coherent and incoherent emission: Frequency filtering then emerged as a tool to reshape the light statistics ~\cite{LopezCarreno2018,Hanschke2020,Phillips2020}. Differently, the modification of the light statistics which we have reported here stems from the interference of the (coherent) radiation from several emitters, thus relying on the spatial features of the radiation rather than the spectral ones. Note that atom-like color center in materials \cite{2016_Juan_Volz_natphys}, trapped ions \cite{2020_Wolf_Schmidt-Kaler_prl,2024_Singh_Bachelard_arxiv}, cold atoms trapped in optical lattices \cite{2005_Bloch_natphys, 2013_Martin_Ye_sci, 2009_Bakr_Greiner_nat} or tweezers \cite{2007_Beugnon_Grangier_natphys,2018_Norcia_Kaufman_prx} are potential platforms to tune the interplay between coherent and incoherent scattering, since they offer remarkable suppression of motional effects and thus allow for high visibility of interference fringes.

The control parameter to achieve extraordinary photon statistics is the number of excitations in the system, $sN$. Only if it is small, strong values of anti- and superbunching can be achieved, i.e.,  in the regime of low photon flux. This demonstrates the appearance of additional mechanisms to control the light statistics from an ensemble of emitters. While the frequency filtering previously used for single emitters could also be employed, it actually leads to lower fluxes. A possible solution could be to manipulate the spatial features of spontaneous emission, e.g., by bringing the emitters closer together. The electromagnetic vacuum modes to which the atoms couple to spontaneously decay then become a shared reservoir for the emitters~\cite{Lehmberg1970}, which in turn leads to a directional, collective spontaneous emission. Pioneered by Dicke~\cite{Dicke1954}, this corresponds to the celebrated phenomenon of superradiance and it has been shown to be a powerful tool to manipulate the emission properties from large atomic systems~\cite{Gross1982,Bohnet2012,Wiegner2015,Ferioli2023,2024_Ori_Shahmoon_prxq}.


{\it Acknowledgements---}R.B. and J.v.Z. gratefully acknowledge funding and support by the  Bavarian Academic Center for Latin America (BAYLAT). This work was funded by the Deutsche Forschungsgemeinschaft (DFG, German Research Foundation) -- Project-ID 429529648 -- TRR 306 QuCoLiMa (``Quantum Cooperativity of Light and Matter''). P.P.A. acknowledges funding from CNPq (Grant No. 152050/2024-8). A.C., P.P.A., and R.B. have received the financial support of the S\~ao Paulo Research Foundation (FAPESP) (Grants No.~2022/06449-9, 2023/07463-8, 2021/04861-7, 2022/00209-6, 2023/07100-2, and 2023/03300-7).\\

\bibliography{g2_bib}

\end{document}


\begin{center}
{\large{ {\bf Supplemental Material for: \\ Light Statistics from Large Ensembles of Independent Two-level Emitters: Classical or Non-classical? }}}

\setcounter{page}{1}

\vskip0.5\baselineskip{M. Bojer$^{1}$, A. Cidrim$^{2}$, P. P. Abrantes$^{2,\,3}$, R. Bachelard$^{2}$, and J. von Zanthier$^{1}$}

\vskip0.5\baselineskip{ {\it $^{1}$Friedrich-Alexander-Universität Erlangen-Nürnberg, Quantum Optics and Quantum Information, Staudtstr. 1, 91058 Erlangen, Germany\\
$^{2}$Departamento de F\'isica, Universidade Federal de S\~ao Carlos,
	Rodovia Washington Lu\'is, km 235—SP-310, 13565-905 S\~ao Carlos, SP, Brazil\\
 $^3$Instituto de F\'isica, Universidade Federal do Rio de Janeiro, 21941-972 Rio de Janeiro, RJ, Brazil}}
\end{center}

\appendix

\setcounter{figure}{0}

\section{Classical antennas} 

A classical model of a driven atomic dipole moment can be found in the Lorentz oscillator model, which describes a damped driven harmonic oscillator. The quantum version is a damped driven quantum harmonic oscillator. In resonance and in the interaction picture, the equation of motion is
\begin{align}
    \dot{\hat{\rho}} = \frac{1}{i\hbar}[\hat{H}, \hat{\rho}] - \gamma (\hat{b}^\dagger \hat{b} \hat{\rho} + \hat{\rho} \hat{b}^\dagger \hat{b} - 2 \hat{b} \hat{\rho} \hat{b}^\dagger),
\end{align}
with $\hat{H}=-i\hbar \Omega (\hat{b}^\dagger - \hat{b})$. In the steady state ($\dot{\hat{\rho}} = 0$) and with the ansatz $\hat{\rho}_{\mathrm{ss}}=\ket{\alpha}\bra{\alpha}$, where $\ket{\alpha}$ is a coherent state, we find that the system reaches the coherent state with $\alpha = -\Omega/\gamma$. In the case of $N$ independent antennas, the steady state is simply given by the $N$th tensor product. Therefore, for this classical antenna model, the second-order autocorrelation function simply yields $g^{(2)}_{\mathbf{k}}=1$. Hence, coherent light statistics is obtained from this description. 

Even if the atoms are modelled as two-level emitters rather than oscillating dipoles, in the single-excitation limit, which is often applied for very low saturation parameters (as considered in this paper), one can find a 1:1 correspondence. Indeed, in this limit the coherent state can be approximated by
\begin{align}
    \ket{\alpha}^{\oplus N} \approx e^{-N|\alpha|^2/2} \ket{0,...,0} + e^{-N|\alpha|^2/2} \alpha \sum_{\mu=1}^N \hat{b}_{\mu}^\dagger \ket{0,...,0}\,.
\end{align}
Therefore, the zero-excitation population corresponding to the density matrix entry $\ketbra{0,...,0}{0,...,0}$, the one-zero excitation coherences corresponding to the entries $\hat{b}_{\mu}^\dagger\ketbra{0,...,0}{0,...,0}$, the one-excitation populations corresponding to the entries $\hat{b}_{\mu}^\dagger\ketbra{0,...,0}{0,...,0}\hat{b}_{\mu}$, and the one-one excitation coherences corresponding to the entries $\hat{b}_{\mu}^\dagger\ketbra{0,...,0}{0,...,0}\hat{b}_{\nu}$ ($\mu\neq \nu$) are approximately given by
\begin{align}
    \ketbra{0,...,0}{0,...,0}:& e^{-N|\alpha|^2} \approx 1 - \frac{sN}{2}\,,\\
    \hat{b}_{\mu}^\dagger\ketbra{0,...,0}{0,...,0}:& e^{-N|\alpha|^2} \alpha \approx - \frac{\Omega}{\gamma} = - \frac{\sqrt{s}}{\sqrt{2}}\,,\\
    \hat{b}_{\mu}^\dagger\ketbra{0,...,0}{0,...,0}\hat{b}_{\mu}:& e^{-N|\alpha|^2} |\alpha|^2 \approx \frac{s}{2}\,,\\
    \hat{b}_{\mu}^\dagger\ketbra{0,...,0}{0,...,0}\hat{b}_{\nu}:& e^{-N|\alpha|^2} |\alpha|^2 \approx \frac{s}{2}\,.
\end{align}
On the other hand, for two-level emitters these same density matrix entries can be approximated by
\begin{align}
    \ketbra{g,...,g}{g,...,g}:& \left(\frac{2+s}{2(1+s)}\right)^{N} \approx 1 - \frac{sN}{2}\,,\\
    \hat{\sigma}_{\mu}^+\ketbra{g,...,g}{g,...,g}:& -\left(\frac{2+s}{2(1+s)}\right)^{N-1}\frac{\sqrt{s}}{\sqrt{2}(1+s)} \approx - \frac{\sqrt{s}}{\sqrt{2}}\,,\\
    \hat{\sigma}_{\mu}^+\ketbra{g,...,g}{g,...,g}\hat{\sigma}_{\mu}^-:& \quad
    \left(\frac{2+s}{2(1+s)}\right)^{N-1}\frac{s}{2(1+s)} \approx \frac{s}{2}\,,\\
    \hat{\sigma}_{\mu}^+\ketbra{g,...,g}{g,...,g}\hat{\sigma}_{\nu}^-:&
    \left(\frac{2+s}{2(1+s)}\right)^{N-2}\left(-\frac{\sqrt{s}}{\sqrt{2}(1+s)}\right)^2 \approx \frac{s}{2}\,.
\end{align}
That is, up to the first order in $s$ and restricting the state to only one excitation, the coherent state and the atomic state can be mapped onto each other. From this perspective, one might conclude that for a vanishing number of excitations the ensemble of two-level emitters will exhibit coherent light statistics. However, as we show in the main text, this would be an erroneous conclusion -- this can be explained by the fact that two-photon correlations are not properly described by single-excitation states.

\section{Explicit calculation of the second-order autocorrelation function}

To calculate the second-order autocorrelation function we start by writing the first-order correlation function explicitly:
\begin{align}
    G^{(1)}_{\mathbf{k}} &= \braket{\hat{E}^{(-)}_{\mathbf{k}}\hat{E}^{(+)}_{\mathbf{k}}} = \sum_{\mu,\nu=1}^N e^{i \mathbf{k}.(\mathbf{R}_\mu-\mathbf{R}_\nu)}\braket{\hat{\sigma}^+_\mu \hat{\sigma}^-_\nu} = \sum_{\mu=1}^N (\braket{\hat{\sigma}^+_\mu \hat{\sigma}^-_\mu}-\braket{\hat{\sigma}^+_\mu}\braket{\hat{\sigma}^-_\mu}) + \sum_{\mu,\nu=1}^N e^{i \mathbf{k}.(\mathbf{R}_\mu-\mathbf{R}_\nu)}\braket{\hat{\sigma}^+_\mu}\braket{\hat{\sigma}^-_\nu}\nonumber\\
    &= \frac{s}{2(1+s)}N - \frac{s}{2(1+s)^2}N + \frac{s}{2(1+s)^2}\left|S(\mathbf{k})\right|^2 = \frac{s}{2(1+s)^2}(sN + \left|S(\mathbf{k})\right|^2)\,.
\end{align}
In the case of the unnormalized second-order autocorrelation function
\begin{align}
    G^{(2)}_{\mathbf{k}} &= \braket{\hat{E}^{(-)}_{\mathbf{k}}\hat{E}^{(-)}_{\mathbf{k}}\hat{E}^{(+)}_{\mathbf{k}}\hat{E}^{(+)}_{\mathbf{k}}} = \sum_{\mu_1,\mu_2,\nu_1,\nu_2=1}^N e^{i \mathbf{k}.(\mathbf{R}_{\mu_1}+\mathbf{R}_{\mu_2}-\mathbf{R}_{\nu_1}-\mathbf{R}_{\nu_2})}\braket{\hat{\sigma}^+_{\mu_1}\hat{\sigma}^+_{\mu_2} \hat{\sigma}^-_{\nu_1}\hat{\sigma}^-_{\nu_2}},
\end{align}
we get three different contribution types corresponding to the expectation values $\braket{\hat{\sigma}^+\hat{\sigma}^-}^2$, $\braket{\hat{\sigma}^+\hat{\sigma}^-}\braket{\hat{\sigma}^+}\braket{\hat{\sigma}^-}$, and $(\braket{\hat{\sigma}^+}\braket{\hat{\sigma}^-})^2$. Note that we have neglected to write the atomic indices, since all atoms are in the same steady state. The first expectation value gives
\begin{align}
    2 \sum_{\substack{\mu,\nu=1\\\mu\neq\nu}}^N \braket{\hat{\sigma}^+\hat{\sigma}^-}^2 = 2 \braket{\hat{\sigma}^+\hat{\sigma}^-}^2(N^2-N)\,,
\end{align}
whereas the second expectation value evaluates to
\begin{align}
    4 \sum_{\substack{\mu,\nu,\eta=1\\ \text{mutually different}}}^N e^{i\mathbf{k}.(\mathbf{R}_\nu - \mathbf{R}_\eta)}\braket{\hat{\sigma}^+\hat{\sigma}^-}\braket{\hat{\sigma}^+}\braket{\hat{\sigma}^-}= 4 \braket{\hat{\sigma}^+\hat{\sigma}^-}\braket{\hat{\sigma}^+}\braket{\hat{\sigma}^-} \left[\left|S(\mathbf{k})\right|^2 (N-2)-N^2+2N\right]\,.  
\end{align}
Finally, the calculation of the last expectation value leads to
\begin{align}
    \sum_{\substack{\mu_1,\mu_2,\nu_1,\nu_2=1\\ \text{mutually different}}}^N e^{i\mathbf{k}.(\mathbf{R}_{\mu_1} + \mathbf{R}_{\mu_2} - \mathbf{R}_{\nu_1} - \mathbf{R}_{\nu_2})}\left(\braket{\hat{\sigma}^+}\braket{\hat{\sigma}^-}\right)^2 = \left(\braket{\hat{\sigma}^+}\braket{\hat{\sigma}^-}\right)^2 \left[\left|S(\mathbf{k})\right|^4-S^2(-\mathbf{k})S(2\mathbf{k})-S^2(-\mathbf{k})^*S(2\mathbf{k})^*+\left|S(2\mathbf{k})\right|^2\right.\nonumber\\
    \left.- 4 \left|S(\mathbf{k})\right|^2 (N-2) + 2(N^2-3N)\right]\,.
\end{align}
Now, by summing up all three terms we obtain for the unnormalized second-order autocorrelation function
\begin{align}
    G^{(2)}_{\mathbf{k}} = \frac{s^2}{4(1+s)^4}\left\{\left|S^2(\mathbf{k})-S(2\mathbf{k})\right|^2 + 4s(N-2)\left|S(\mathbf{k})\right|^2 + 2sN\left[2+s(N-1)\right]\right\}\,.
\end{align}
This leads to the following expression for the normalized second-order autocorrelation function:
\begin{align}
    g^{(2)}_{\mathbf{k}} = \frac{G^{(2)}_{\mathbf{k}}}{\left[G^{(1)}_{\mathbf{k}}\right]^2} = \frac{\braket{\hat{E}^{(-)}_\mathbf{k} \hat{E}^{(-)}_\mathbf{k} \hat{E}^{(+)}_\mathbf{k} \hat{E}^{(+)}_\mathbf{k}}}{\braket{\hat{E}^{(-)}_\mathbf{k}\hat{E}^{(+)}_\mathbf{k}}^2}	
    = \frac{2sN\left[2 +s(N-1)\right]+ 4s(N-2)\left|S(\mathbf{k})\right|^2 + \left|S^2(\mathbf{k}) - S(2\mathbf{k})\right|^2}{\left(sN + |S(\mathbf{k})|^2\right)^2}\,.
\end{align}

\section{Ordered case}

In the case of a one-dimensional chain with $\phi= d\,\hat{\mathbf{n}}.(\mathbf{k}_{\text{obs}}-\mathbf{k}_L)$ the structure factor becomes a geometric sum
\begin{align}
    S(\mathbf{k}) = \sum_{\mu=1}^{N} e^{i\mu \phi} = e^{i\phi} \frac{1-e^{iN\phi}}{1-e^{i\phi}} = \frac{1-e^{iN\phi}}{e^{-i\phi}-1}\,.
\end{align}

{\em Superbunching.---} The superbunching condition corresponds to a fully destructive interference, i.e., $S(\mathbf{k})=0$ implying $(e^{i\phi})^N = 1$. That is, $e^{i\phi}$ needs to be an $N$th root of unity. However, this also implies that 
\begin{align}
    S(2\mathbf{k}) = \frac{1-e^{2iN\phi}}{e^{-2i\phi}-1} = \frac{1-(e^{iN\phi})^2}{e^{-2i\phi}-1} = 0\,.
\end{align}
An exception to this is the particular case of an even number of atoms and an optical path of $\phi=(2m+1)\pi$ with $m \in \mathbb{Z}$. In this case, the structure factor reads
\begin{align}
    S(\mathbf{k}) = \sum_{\mu=1}^{N} e^{i\mu \pi} = \sum_{\mu=1}^{N} (-1)^{\mu}\,.
\end{align}
Since $N$ is even, $S(\mathbf{k})=0$, but $S(2\mathbf{k})=N$. Therefore, in this particular case, in the limit $sN\rightarrow 0$ the second-order autocorrelation function scales as $g^{(2)}_{\mathbf{k}}\propto 1/s^2$.\\

\begin{figure}
\centering
\includegraphics[width=0.5\textwidth]{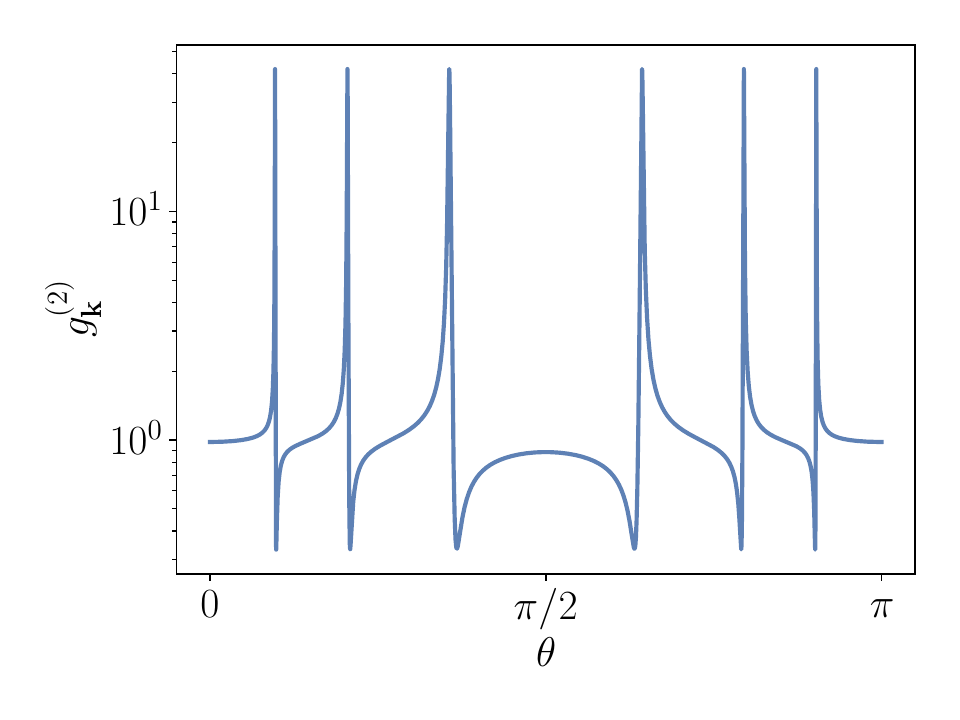}
 \caption{Second-order autocorrelation function as a function of the polar angle $\theta$ (measured from the $z$-axis) for $s=10^{-3}$ and $N=100$. The atoms are located along the $x$-axis, i.e., $\hat{\mathbf{n}}=\hat{\mathbf{x}}$, the laser is parallel to the $z$-axis, i.e., $\mathbf{k}_L \parallel \hat{\mathbf{z}}$ and the detector is rotated in the $xz$-plane, so that $\phi = d k \sin\theta$. As can be seen, both strong super- and antibunching can be observed depending on the value of $\theta$, i.e, the observation direction $\mathbf{k}_{\text{obs}}$.}
\label{Fig:g2_phi}
\end{figure}

{\em Antibunching.---} Achieving  antibunching requires minimizing the last term in the numerator of $g^{(2)}_\mathbf{k}$, that is, $S^2(\mathbf{k})-S(2\mathbf{k})=0$. In this case, consider $(e^{i\phi})^{N-1}=1$, i.e., an $(N-1)$th root of unity. Then, calculating the structure factor leads to
\begin{align}
    S(\mathbf{k}) = \frac{1-e^{iN\phi}}{e^{-i\phi}-1} = \frac{1-e^{i(N-1)\phi}e^{i\phi}}{e^{-i\phi}-1} = \frac{1-e^{i\phi}}{e^{-i\phi}-1} = e^{i\phi}
\end{align}
and thus $S(2\mathbf{k})=e^{2i\phi}$. Hence, we have $S^2(\mathbf{k})-S(2\mathbf{k})=0$, but $S(\mathbf{k})\neq 0$.\\
An example of angular pattern with the scalings discussed in the main text is presented in Fig.~\ref{Fig:g2_phi}.

\section{Disordered case}

In this section we discuss the numerical analysis of the structure factor expressions that arise for the conditions of super- and antibunching. 

\begin{figure}[h!]
\centering
\includegraphics[width=0.45\textwidth]{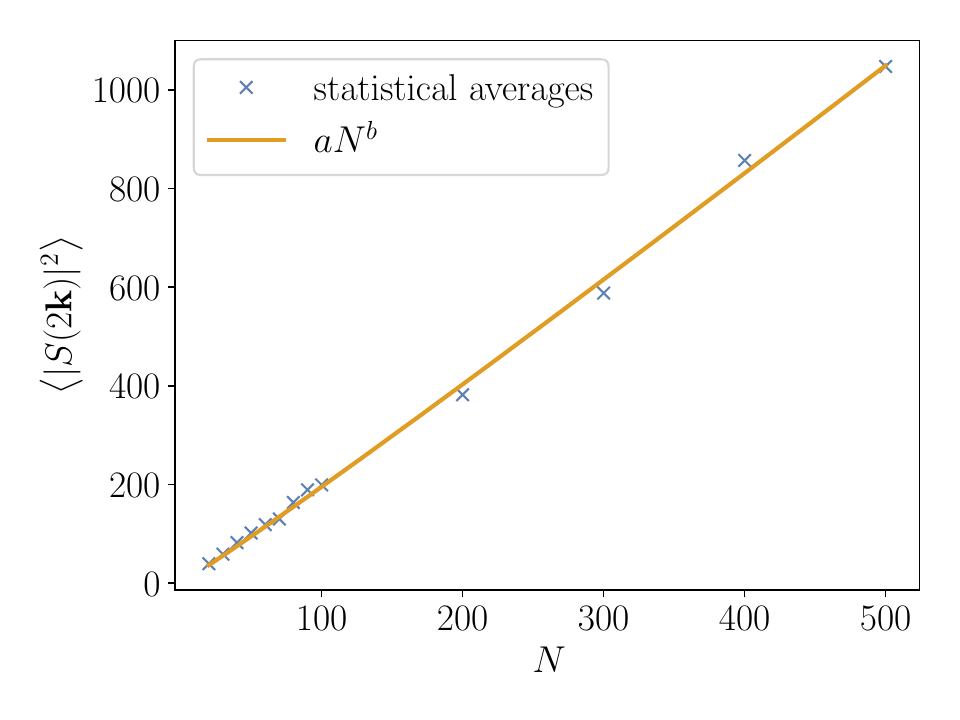}\includegraphics[width=0.45\textwidth]{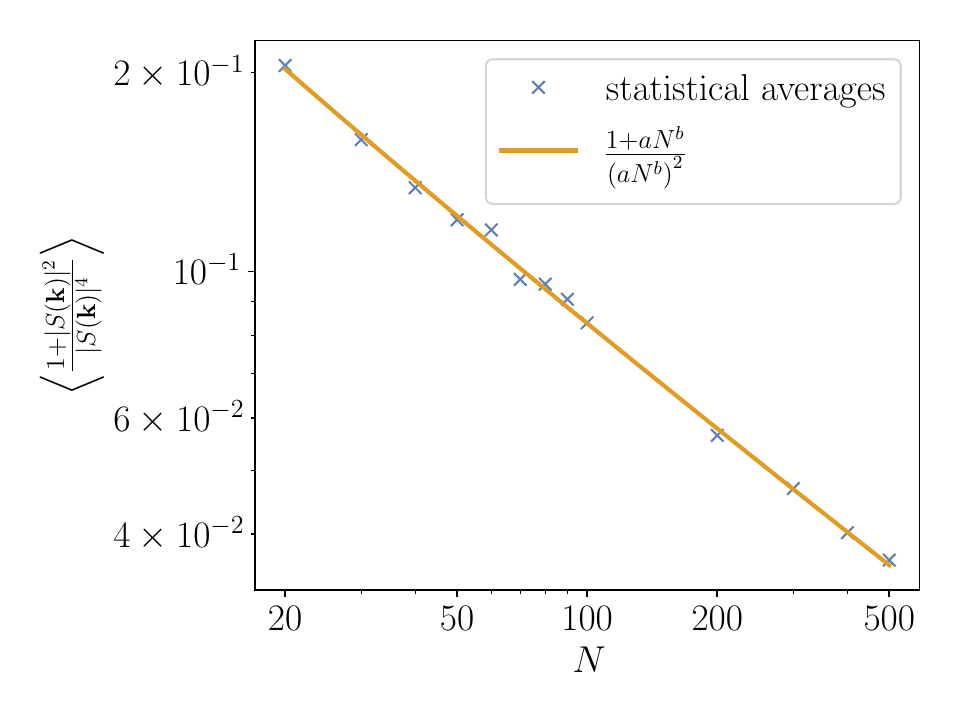}
\caption{Left: Analysis of the structure factor expression $\left|S(2\mathbf{k})\right|^2$ under the condition of $S(\mathbf{k})=0$. The fit of the ensemble averages reveals $b\approx 1$. Right: Analysis of the structure factor expression $(1+\left|S(\mathbf{k})\right|^2)/\left|S(\mathbf{k})\right|^4$ under the condition of $S^2(\mathbf{k})-S(2\mathbf{k})=0$. The fit of the ensemble averages reveals $b\approx 0.5$.}
\label{Fig:S2k_fit}
\end{figure}

In the case of superbunching, we need to evaluate the expression $\left|S(2\mathbf{k})\right|^2$ under the condition of $S(\mathbf{k})=0$. Therefore, we calculated for several numbers of atoms the ensemble averages over $200$ realizations and determined the scaling with $N$ via a fit of the form $aN^b$ (see Fig.~\ref{Fig:S2k_fit}, left panel). The fit shows a scaling of $b\approx 1$.

In the case of antibunching, we need to evaluate the expression $(1+\left|S(\mathbf{k})\right|^2)/\left|S(\mathbf{k})\right|^4$ under the condition of $S^2(\mathbf{k})-S(2\mathbf{k})=0$. Therefore, we calculated for several numbers of atoms the ensemble averages over $200$ realizations and determined the effective scaling with $N$ via a fit of the form $(1+aN^b)/(aN^b)^2$ (see Fig.~\ref{Fig:S2k_fit}, right panel). The fit exhibits a scaling of $b\approx 0.5$.


\section{Conditions for generalized ``superbunching" and ``antibunching"}

In the following section, we generalize the conditions for ``superbunching" and ``antibunching" for an arbitrary correlation order $m$. As we show, the condition for generalized superbunching is still a strong destructive first-order interference, expressed by $S(\mathbf{k})=0$. In contrast, in the case of generalized antibunching, the mathematical condition becomes more involved.\\
We start by considering the symmetric permutation group $S_m$. The number of conjugacy classes of $S_m$ equals the number of integer partitions of $m$~\cite{dummit2003abstract}. Let $P_{c_1,...,c_m}\vdash m$ denote an integer partition of $m$ with $m=\sum_{j=1}^{m} c_j j$ and $c_j \in \{0,...,m\}$. Thereby, the partition $P_{c_1,...,c_m}\vdash m$ characterized by $c_1,...,c_m$ corresponds to the conjugacy class of an element $\sigma \in S_m$ with $c_j$ disjoint cycles of length $j$. According to the orbit-stabilizer theorem, the cardinality $C_{c_1,...,c_m}$, i.e., the number of elements of the conjugacy class of an element $\sigma$, is equal to the index $|S_m:\mathrm{C}_{S_m}(\sigma)|=|S_m|/|\mathrm{C}_{S_m}(\sigma)|$ of the centraliser $\mathrm{C}_{S_m}(\sigma)$ of $\sigma$. The number of elements in the centraliser of the permutation $\sigma$ with cycle characterization $c_1,...,c_m$ can be obtained by counting permuting permutations and is given by~\cite{dummit2003abstract}
\begin{align}
	|\mathrm{C}_{S_m}(\sigma)| = \prod_{j=1}^{m} c_j! j^{c_j}\,.
\end{align}
Thus, the number of elements in the corresponding conjugacy class is~\cite{dummit2003abstract}
\begin{align}
    C_{c_1,...,c_m} = \frac{m!}{\prod_{j=1}^{m} c_j! j^{c_j}}\,.
\end{align}
We now apply the concept of conjugacy classes to the calculation of the $m$th-order autocorrelation function. Therefore, let us first consider the sum
\begin{align}
    \sum_{\substack{\mu_1,...,\mu_m=1\\ \text{mutually different}}}^{N} 1 = m! \binom{N}{m} = N(N-1)...(N-m+1)\,.
\end{align}          
The last expression is known as a falling factorial and gives a polynomial in $N$
\begin{align}
    \label{eq:stirling}
    N(N-1)...(N-m+1) = \sum_{j=1}^{m} s(m,j) N^j\,,
\end{align}
where the coefficients $s(m,j)$ are the Stirling numbers of the first kind. They can be written as $s(m,j)=(-1)^{m-j}c(m,j)$ with $c(m,j)$ being the unsigned Stirling numbers of the first kind~\cite{Stanley_2011}. It is well-known that $c(m,j)$ gives the number of permutations within the permutation group $S_m$, which consist of exactly $j$ many cycles. Therefore, the unsigned Stirling numbers of the first kind $c(m,j)$ are connected to the cardinalities $C_{c_1,...,c_m}$ via
\begin{align}
    c(m,j) = \sum_{\substack{c_1,...,c_m \\ \sum_{l} c_l = j}} C_{c_1,...,c_m}\,.
\end{align}
This allows us to write the sum in Eq.~\eqref{eq:stirling} with respect to the cardinalities of the conjugacy classes of $S_m$ as
\begin{align}
    \label{eq:partitions}
	\sum_{j=1}^{m} s(m,j) N^j =C_{m,0,...,0} N^m - C_{m-2,1,0,...,0} N^{m-1}+ C_{m-3,2,0,...,0} N^{m-2} + C_{m-3,1,1,0,...,0} N^{m-2} -...\,.
\end{align}
Now, let us define
\begin{align}
	\prod_{\nu_1 \in \{ \mu_{2},...,\mu_{m}\}}\!\!\!(1-\delta_{\mu_1,\nu_1}) \times \!\!\!\!\!\!\prod_{\nu_2 \in \{ \mu_{3},...,\mu_{m}\}}\!\!\!(1-\delta_{\mu_2,\nu_2}) \times ... \times (1-\delta_{\mu_{m-1},\mu_m})\eqqcolon 1 + f(\delta_{\mu_1,\mu_2},...,\delta_{\mu_1,\mu_m},\delta_{\mu_2,\mu_3},...,\delta_{\mu_2,\mu_m},...,\delta_{\mu_{m-1},\mu_m})\,,
\end{align}
where $f(\delta_{\mu_1,\mu_2},...,\delta_{\mu_1,\mu_m},\delta_{\mu_2,\mu_3},...,\delta_{\mu_2,\mu_m},...,\delta_{\mu_{m-1},\mu_m})$ is a multivariate polynomial of degree $m-1$. Then, we can write
\begin{align}
	\sum_{\substack{\mu_1,...,\mu_m=1\\ \text{mutually different}}}^{N} 1 = \sum_{j=1}^{m} s(m,j) N^j = \sum_{\mu_1,...,\mu_m=1}^{N} [1 + f]\,.
\end{align}
The last expression allows us to conclude that the $s(m,m-1)$ term (counting transpositions) comes from the single delta contributions in $f$, the $s(m,m-2)$ term (counting double transpositions and 3-cycles) comes from the double delta contributions in $f$, and so on. In summary, a Kronecker delta $\delta_{i,j}$ can be associated with the transposition $(i j)$. Therefore, the following sum
\begin{align}
	\sum_{\substack{\mu_1,...,\mu_m=1\\ \text{mutually different}}}^{N} h(\mu_1,...,\mu_m)
\end{align}
with an arbitrary symmetric function $h(\mu_1,...,\mu_m)$ with respect to $\mu_1,...,\mu_m$ can be evaluated via the decomposition in partitions, as in Eq.~\eqref{eq:partitions}.\\
With this at hand, let us determine the conditions for generalized superbunching and antibunching via a Taylor expansion of the $m$th-order autocorrelation function in the saturation parameter $s$ reading
\begin{align}
	g^{(m)}_{\mathbf{k}} = \tilde{g}^{(m)}_{\mathbf{k},0} + \tilde{g}^{(m)}_{\mathbf{k},1} s + \mathcal{O}(s^2)\,.
\end{align}
The zeroth order is obtained for $s=0 \Leftrightarrow \braket{\hat{\sigma}^+\hat{\sigma}^-} = \braket{\hat{\sigma}^+}\braket{\hat{\sigma}^-}$, where we neglected the index $\mu$ since all atoms are in the same state. Then,
\begin{align}
	\tilde{G}^{(m)}_{\mathbf{k},0} = (\braket{\hat{\sigma}^+}\braket{\hat{\sigma}^-})^m |S^{(m)}(\mathbf{k})|^2 = \frac{s^m}{2^m (1+s)^{2m}} |S^{(m)}(\mathbf{k})|^2\,,
\end{align}
where we defined the generalized $m$th-order structure factor
\begin{align}
	S^{(m)}(\mathbf{k}) \coloneqq \sum_{\substack{\mu_1,...,\mu_m=1\\ \text{mutually different}}}^{N} e^{i\mathbf{k}\mathbf{R}_{\mu_1}}...e^{i\mathbf{k}\mathbf{R}_{\mu_m}}\,.
\end{align}
Furthermore, we have
\begin{align}
    &G^{(1)}_{\mathbf{k}} = \frac{s}{2(1+s)^2}[sN+|S(\mathbf{k})|^2]\nonumber\\
    &\Rightarrow \left[G^{(1)}_{\mathbf{k}}\right]^m = \frac{s^m}{2^m (1+s)^{2m}}[sN + |S(\mathbf{k})|^2]^m
\end{align}
and thus
\begin{align}
	\frac{\tilde{G}^{(m)}_{\mathbf{k},0}}{\left[G^{(1)}_{\mathbf{k}}\right]^m} = \frac{|S^{(m)}(\mathbf{k})|^2}{[sN+|S(\mathbf{k})|^2]^m}.
\end{align}
Therefore, if we have a strong destructive first-order interference, i.e., $S(\mathbf{k})=0$, then
\begin{align}
	\tilde{g}^{(m)}_{\mathbf{k},0} = \frac{|S^{(m)}(\mathbf{k})|^2_{|S(\mathbf{k})=0}}{(sN)^m} \propto \frac{1}{s^m}\,,
\end{align}
such that a strong ``superbunching" can be achieved. In contrast, if $S^{(m)}(\mathbf{k})=0$, we have $\tilde{g}^{(m)}_{\mathbf{k},0}=0$ and $g^{(m)}_{\mathbf{k}}\propto s$, such that a strong ``antibunching" is obtained. To finish our results, we explicitly calculate $S^{(m)}(\mathbf{k})$. According to the discussion about the decomposition in partitions and conjugacy classes, we find
\begin{align}
	S^{(m)}(\mathbf{k}) = \sum_{P_{c_1,...,c_m}\vdash m} (-1)^{m-\sum_{j=1}^{m} c_j} m! \prod_{l=1}^{m} \frac{S^{c_l}(l \mathbf{k})}{c_l! l^{c_l}}\,.
\end{align}
To conclude this section, let us present some simple examples ($m=2$, $m=3$, and $m=4$). In the case of $m=2$, we have two integer partitions, namely $P_{2,0}=1+1$ and $P_{0,1}=2$. Therefore,
\begin{align}
    S^{(2)}(\mathbf{k}) = S^2(\mathbf{k}) - S(2\mathbf{k})\,.
\end{align}
In the case of $m=3$, we have three integer partitions, namely $P_{3,0,0}=1+1+1$, $P_{1,1,0}=1+2$, and $P_{0,0,1}=3$. Thus,
\begin{align}
	S^{(3)}(\mathbf{k}) = S^3(\mathbf{k}) - 3 S(\mathbf{k})S(2\mathbf{k}) + 2 S(3\mathbf{k})\,.
\end{align}
Finally, in the case of $m=4$, we have five integer partitions, namely $P_{4,0,0,0}=1+1+1+1$, $P_{2,1,0,0}=1+1+2$, $P_{0,2,0,0}=2+2$, $P_{1,0,1,0}=1+3$, and $P_{0,0,0,1}=4$. Then,
\begin{align}
    S^{(4)}(\mathbf{k}) = S^4(\mathbf{k}) - 6 S^2(\mathbf{k}) S(2\mathbf{k}) + 3 S^2(2\mathbf{k})+ 8 S(\mathbf{k})S(3\mathbf{k}) - 6 S(4\mathbf{k})\,.
\end{align}
\bibliography{supp_bib}